\documentclass[fleqn,usenatbib]{mnras}

\usepackage{newtxtext,newtxmath}

\usepackage[T1]{fontenc}
\usepackage{ae,aecompl}
\usepackage{soul}

\usepackage{graphicx}	
\usepackage{amsmath}	
\usepackage[usenames]{xcolor}
\usepackage{natbib}
\usepackage{hyperref}
\usepackage{xfrac}
\usepackage{verbatim}

\def\equationautorefname~#1\null{equation~(#1)}

\DeclareMathAlphabet{\mathpzc}{OT1}{pzc}{m}{it}\definecolor{purple}{RGB}{160,32,240}

\newcommand{\rev}[1]{\textcolor{black}{#1}}
\newcommand{\reva}[1]{\textcolor{black}{#1}}
\newcommand{\revb}[1]{\textcolor{black}{#1}}

\newcommand{\eugene}[1]{\textcolor{black}{#1}}

\newcommand{\Msun}{M_{\odot}}

\newcommand{\gcc}{\rm g/cm^3}

\newcommand{\rhon}{\rho_{\rm neb}}

\newcommand{\sch}{s_{\rm c}}
\newcommand{\Tc}{T_{\rm c}}

\newcommand{\rhos}{\rho_{\rm solid}}

\title[\rev{Humpty Dumpty}]{\rev{Chondrules from high-velocity collisions: thermal histories and the agglomeration problem}}

\author[Choksi et al.]{Nick Choksi$^{1 \href{https://orcid.org/0000-0003-0690-1056}{\includegraphics[scale=0.4]{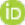}}}$\thanks{E-mail: nchoksi@berkeley.edu},
Eugene Chiang$^{1,2\href{https://orcid.org/0000-0002-6246-2310}{\includegraphics[scale=0.4]{FIGURES/orcid.pdf}}}$, Harold C. Connolly Jr.$^{3}$, Zack Gainsforth$^{4}$, \and Andrew J. Westphal$^{4}$
\\
$^{1}$Astronomy Department, Theoretical Astrophysics Center, and Center for Integrative Planetary Science, University of California, Berkeley, CA \\ 
$^{2}$Department of Earth and Planetary Science, University of California, Berkeley, CA \\ 
$^{3}$Department of Geology, School of Earth and Environment, Rowan University, Glassboro, NJ \\ 
$^{4}$Space Sciences Laboratory, University of California, Berkeley, CA \\
}

\date{Released \today}

\pubyear{2020}

\begin{document}
\label{firstpage}
\pagerange{\pageref{firstpage}--\pageref{lastpage}}
\maketitle

\begin{abstract}
\rev{We assess whether chondrules, once-molten mm-sized spheres filling the oldest meteorites, could have formed from super-km/s collisions between planetesimals in the solar nebula. High-velocity collisions release hot and dense clouds of silicate vapor which entrain and heat chondrule precursors. \reva{ \revb{Thermal histories of CB chondrules are reproduced for colliding bodies $\sim$10--100 km in radius. The slower cooling rates of non-CB, porphyritic chondrules point to colliders with radii $\gtrsim$ 500 km. How chondrules,} collisionally dispersed into the nebula, agglomerated into meteorite parent bodies remains a mystery. The same orbital eccentricities and inclinations that enable energetic collisions prevent planetesimals from re-accreting chondrules efficiently and without damage; thus the sedimentary laminations of the CB/CH chondrite Isheyevo are hard to explain by direct fallback of collisional ejecta.} At the same time, planetesimal surfaces may be littered with the shattered remains of chondrules. The micron-sized igneous particles recovered from comet 81P/Wild-2 may have originated from in-situ collisions and subsequent accretion in the proto-Kuiper belt, obviating the need to transport igneous solids across the nebula. Asteroid sample returns from \textit{Hayabusa2} and \textit{OSIRIS-REx} may similarly contain chondrule fragments. }
\end{abstract}



\section{Introduction}
\label{sec:Intro}
Beneath 
the fusion crusts of the
most primitive stony meteorites
lies a profusion of millimeter-sized
igneous spheres. 
These chondrules, which can fill $\gtrsim$ 50\% of  meteorite volumes, are among the oldest creations of the solar system, with lead isotopic ages of 4.562--4.567 billion years, and near-solar compositions in refractory elements. 
Petrologic studies indicate 
that chondrules were heated to melting
temperatures for a few minutes, and 
cooled over hours to days. 
Their roundness 
implies that chondrules were melted while suspended in space,
so that surface
tension pulled their shapes
into spheres. For reviews, see \cite{connolly_jones_review}
and \cite{russell_etal_2018}.

From the petrologic 
data we can infer some rough
orders of magnitude characterizing
the chondrule formation
environment. The fact that chondrules were at least partially
molten implies an ambient temperature 
of $T \sim 2000$ K.
A single chondrule radiating
into vacuum at this temperature
would cool off unacceptably
fast, within seconds;
therefore chondrules
must have been immersed in, and
kept warm by, a gas
of high heat capacity, or a
radiation bath maintained
by an optically thick medium,
or both. The former possibility
is further supported by the retention of volatile
elements -- principally sodium -- within some 
chondrules, 
requiring ambient Na partial
pressures of order $\sim$$10^{-3}$ bars while $T \sim 2000$ K \citep{alexander_etal_2008, fedkin_grossman_2013}. Gas at 2000 K 
has a sound speed 
ranging from
$c_{\rm s} \sim 0.7$ to 3
km/s, depending on whether
it is composed predominantly of silicates or hydrogen.
A characteristic length scale
for the formation environment 
is given by the sound speed multiplied by the cooling time, 
$R \sim 4 \times 10^4 \, {\rm km}\, (c_{\rm s}/{\rm km} \, {\rm s}^{-1}) (t_{\rm cool}/10 \, {\rm hr})$. 

These scales, which describe
a formation setting that was hot,
pressurized, and compact, do not
recall those of the solar nebula
(a.k.a. the protoplanetary disc), 
which for the most part was cold ($\lesssim 300$ K), rarefied ($\lesssim 10^{-4}$ bar in hydrogen, and orders of magnitude less in other elements), and
extensive 
(with disc scale heights $\gtrsim 10^{7}$ km; e.g.  \citealt{williams_cieza_2011};
\citealt{armitage_2011}). This mismatch
argues against purely nebular processes for creating chondrules \citep[e.g.][]{desch_connolly_2002}.
Furthermore, the hydrogen-rich composition of the nebula does not provide the high oxygen fugacities (oxygen partial pressures) required to form the iron-rich silicates present in chondrules \citep[e.g.][]{ebel_grossman_2000, grossman_etal_2008}.

The comparatively small, fast, and energetic scales inferred from chondrule petrology may instead be compatible 
with collisions between planetesimals. \rev{Bodies on heliocentric orbits with eccentricities $e$ and mutual inclinations $i$ collide with relative velocities $u_{\rm rel} \sim \sqrt{e^2 + i^2}u_{\rm K}$, where $u_{\rm K}$ is the Keplerian orbital velocity ($u_{\rm K} \sim 20$ km/s in the main asteroid belt at 3 au).} Impacts at $u_{\rm rel} \sim \mathcal{O}(1\,\rm km/s)$ heat rock to a temperature of $T \sim \mu m_{\rm p} u_{\rm rel}^2/k \sim \mathcal{O} (10^3 \,{\rm K})$, where $\mu \sim 30$ is the mean molecular weight of 
silicate gas, $m_{\rm p}$ is the proton mass, and $k$ is Boltzmann's constant. The durations of heating and cooling
should scale with the sizes of the colliding planetesimals. For example, if the colliding bodies are $R_{\rm pl} \sim \mathcal{O}(100 \, {\rm km})$ in size, we might expect the initial fireball to last the ``smash-through'' time of $R_{\rm pl} / u_{\rm rel} \sim \mathcal{O}(10^2 \, {\rm s})$, consistent with a minutes-long
heating event.
The collisional destruction
of the bodies releases 
an expanding cloud of 
debris and vapor,
the thermodynamics of which could
conceivably reproduce chondrule
cooling rates. Fleshing out
this possibility with a
quantitative model is a goal of this paper.

Petrologic support for a collisional origin has been building, \reva{especially for CB/CH chondrites, whose metal nodules (a.k.a. blebs), lack of fine-grained matrix, and  skeletal, non-porphyritic chondrule textures have been interpreted as signatures of melt production and vapor condensation from a hypervelocity impact, i.e. one fast enough to shock compress solids and produce melt and vapor (e.g. \citealt{campbell_etal_2002,krot_etal_2007, ivanova_etal_2008}).} \cite{krot_etal_2005} found that CB chondrules crystallized 4--5 Myr after the formation of the oldest objects in the solar system (calcium-aluminium-rich inclusions in CV chondrites) and argued that by this time, the solar nebula may have largely dissipated, ruling out a purely nebular origin. 
\reva{\cite{fedkin_etal_2015} reproduced the elemental profiles of metal grains and chondrules in CB chondrites by modeling the condensation of an impact plume from a differentiated, water-rich asteroid.}

\cite{asphaug_etal_2011} pioneered numerical simulations of
chondrule formation in planetesimal collisions, considering impact
speeds $u_{\rm rel} \sim \mathcal{O}(100\,\rm m/s)$, comparable to the surface escape velocities of $R_{\rm pl} \sim 100$ km asteroids. Collisions at these relatively low velocities do not melt solids; chondrules were instead envisioned to be liquid droplets sprayed out from planetesimal interiors made molten by $^{26}$Al. \cite{sanders_scott_2012} argued that this scenario could resolve many of the petrologic difficulties of non-collisional formation models. \reva{Taking \cite{asphaug_etal_2011} as a starting point, \cite{hewins_etal_2018} numerically simulated the hydrodynamic and thermal evolution of a melt ejecta plume, finding thermal histories of plume gas parcels in good agreement with CB chondrule thermal histories determined experimentally.}

A serious and still unsolved problem with colliding planetesimals with molten interiors is that such bodies are prone to chemically differentiate and produce droplets with non-solar compositions \citep[][]{lichtenberg_etal_2018}. \cite{johnson_melosh_2015} circumvented this difficulty
by using higher velocity impacts to melt solar-composition solids directly; they found that at $u_{\rm rel} \gtrsim 2.5$ km/s, melt is created and jetted out of the impact site. Droplet sizes were estimated to be on the order of 1--10 mm for impactor diameters of 100--1000 km (see also \citealt{johnson_melosh_2014}). For these same parameters, chondrule cooling rates of 100--3000 K/hr, compatible with experimental petrologic constraints, were inferred from radiative transfer modeling of the optically thick ejecta. In the impact simulations of \citet{johnson_melosh_2015} 
and \citet{wakita_etal_2017,wakita_etal_2021}, only a small fraction of the initially solid colliding mass, up to 7 percent, is jetted out as melt. Whether this efficiency of melt production is adequate to re-process enough of the main asteroid
belt into the chondrites sampled on Earth is unclear.

Here we further explore a collisional genesis for chondrules. New developments in modeling shocked forsterite and silica demonstrate that vapor can be produced in abundance through hypervelocity impacts between planetesimals which are initially solid, at least in their outer layers  \citep{kraus_etal_2012, carter_stewart_2020, davies_etal_2020}. \cite{stewart_lpsc7} suggested that the resulting hot silicate vapor cloud (a.k.a vapor plume) may be conducive to chondrule formation. We investigate this possibility by studying the hydrodynamic and thermodynamic evolution of 
the cloud, including its interaction with the hydrogen of the solar nebula. By contrast to \citet{johnson_melosh_2015} and related studies, we do not rely on melt created at the impact site to form chondrules. Actually, we do not model the impact dynamics at all. Instead we begin our analysis post-impact, assuming that a cloud of hot vapor has been released from the collision, and that this vapor contains initially solid or liquid debris. 
We study the subsequent evolution of this debris, including how it is heated and cooled by ambient vapor and radiation.

Our treatment of the vapor cloud and of the condensed particles embedded within it is zero-dimensional: we do not spatially resolve the cloud, but restrict our attention to its mean properties, e.g. density and temperature, and their evolution with time. Our model is similar in quantitative detail and complementary in scope to that of \cite{dullemond_etal_2014,dullemond_etal_2016}; whereas they considered the evolution of a cloud containing essentially only
molten chondrules, and no vapor except what outgasses from the melt, we study the converse problem of a pure vapor cloud in which solid/liquid particles are sparsely embedded. While the order-of-magnitude style of our approach precludes us from examining the complicated dynamics of the collision itself (cf. \citealt{johnson_melosh_2015}; \citealt{wakita_etal_2017,wakita_etal_2021}; \citealt{davies_etal_2020}),  
it does allow us to describe, broadly and intuitively, the simpler post-collision evolution of the silicate cloud, and to economically survey a range of possible outcomes. Our semi-analytic approach thus complements that of \cite{stewart_lpsc6}, who presented an ab initio 3D smoothed-particle-hydrodynamics (SPH) numerical simulation of one vaporizing collision, \reva{and of \cite{hewins_etal_2018}, whose 3D adaptive-mesh simulation of a melt ejecta plume was initialized using the SPH impact model of \cite{asphaug_etal_2011}.}
While our fiducial parameters will be for the main asteroid belt, we will also scale our results to the proto-Kuiper belt, to see if we might also reproduce the  chondrule-like thermal histories inferred from cometary samples returned by the {\it Stardust} mission \citep{nakamura_etal_2008, jacob_etal_2009, bridges_etal_2012, gainsforth_etal_2015}.

Just as important as the heating and cooling processes is the mechanism by which chondrules agglomerated into meteorite parent bodies. However they were melted, chondrules seem to have assembled into larger bodies with remarkable efficiency, as chondrites make up $\sim$80\% of meteorite falls (\citealt{russell_etal_2018}),\footnote{Modulo the bias of meteorite collections toward objects with sufficient material strength to survive the fall.} and chondrules fill the majority of chondrite volumes (\citealt{weisberg_etal_2006}).
\reva{Agglomeration must have occurred at velocities low enough to avoid shattering chondrules, no more so than in the CB/CH chondrite Isheyevo whose laminations  point to gentle, sedimentary layering of size-sorted material \citep{garvie_etal_2017}.}
Some have suggested that chondrules agglomerate immediately after a planetesimal collision, either re-accreting as fallout over days to weeks \citep{asphaug_etal_2011, morris_etal_2015}, or collecting in local, potentially self-gravitating overdensities \citep{stewart_lpsc3}. Others posit that chondrules are dispersed into the solar nebula and gradually accreted onto planetesimals over the disk lifetime \citep{johansen_etal_2015, hasegawa_etal_2016a, hasegawa_etal_2016b}. 
We will address the question of agglomeration in the context of high-velocity collisions between planetesimals. 

The rest of this paper is organized as follows. In section \ref{sec:model} we describe our model of chondrule heating and cooling using a fiducial set of initial parameters appropriate to a vaporizing collision in the main asteroid belt of the solar nebula. There we also explore how our results vary over the space of initial cloud properties (density, temperature, size) and protoplanetary disc conditions, including those that might have characterized the proto-Kuiper belt. \rev{Section \ref{sec:agglom} tests several hypotheses for how chondrules might have agglomerated after their creation from high-speed collisions.} A summary and outlook are given in section \ref{sec:conclusions}.

\rev{
\section{Vapor Cloud and Chondrule Thermodynamics}
\label{sec:model}
}

\begin{figure*}
\includegraphics[width=\textwidth]{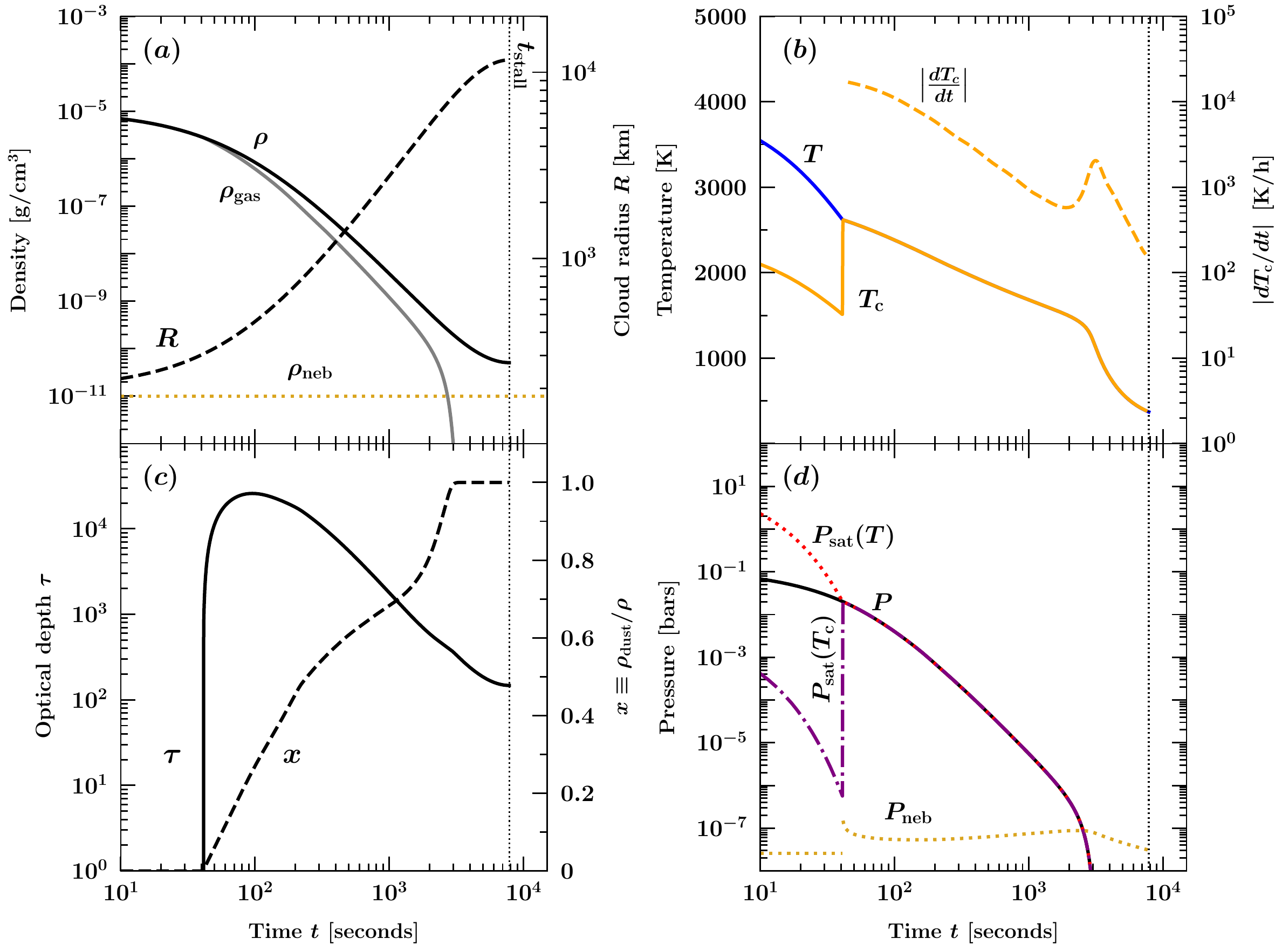}
\caption{Evolution of the silicate vapor cloud and an embedded chondrule precursor of radius $\sch = 0.3$ mm.
The cloud expands
freely at first ($R\propto t$) and its total density $\rho$ scales as $1/R^3$ (panel a). The cloud temperature $T$ and pressure $P$ initially fall adiabatically (panels b and d), while the proto-chondrule is heated conductively by the silicate vapor and cools by blackbody emission to reach an equilibrium temperature $T_{\rm c}< T$  (panel b). The chondrule precursor does not vaporize because its saturation vapor pressure $P_{\rm sat}(T_{\rm c}) < P$ (panel d). At $t \sim 40$ s, the cloud saturates and vapor starts to condense into dust of mass fraction $x = \rho_{\rm dust}/\rho$ (panel c); the gas density $\rho_{\rm gas}$ is now less than $\rho$ (panel a). Once saturated, the cloud stays saturated with $P = P_{\rm sat}(T)$, and for a time, $T$ falls more slowly than along the original adiabat because of latent heat release. The nebular pressure $P_{\rm neb}$ just outside the cloud is higher than before dust condensation because of heating by dust-emitted radiation. The dust optical depth $\tau$ is always $\gg 1$ (panel c) but by $t \sim 3000$ s has decreased enough that radiative cooling causes $T$ to drop faster, all of the remaining vapor to condense ($x=1$), and $P$ to nosedive. Radiative heating by dust enforces $T_{\rm c} = T$; the chondrule cools along with the cloud over a couple hours (see also $dT_{\rm c}/dt$ in panel b). \rev{We end our calculation at $t_{\rm stall} \sim 8000 \, {\rm s} \sim 2.2$ hours, when the nebular headwind has halted the cloud's expansion and ambient hydrogen starts to backfill the pressure-less cavity left by condensation.}}
\label{fig:background}
\end{figure*}


\reva{At a heliocentric distance of $\sim$3 au where the main asteroid belt resides, collisions fast enough to vaporize rock, with $u_{\rm rel} \gtrsim 3$ km/s, implicate crossing orbits having eccentricities and inclinations $e,i \gtrsim 0.1$. We stage our calculations at a time during the asteroid belt's past when the required ellipticities and inclinations have developed as a result of gravitational scatterings off large bodies (e.g. \citealt{raymond_nesvorny_2020}; \citealt{carter_stewart_2020}). During this era, dynamical heating competes effectively against drag exerted by residual nebular gas to keep at least the largest asteroids stirred. Thus we consider the collision of two planetesimals moving on elliptical and/or inclined orbits,
and further assume the bodies to have roughly equal masses for simplicity.}

\reva{Given our set-up, collisional debris, including silicate} \reva{vapor, is ejected onto orbits having eccentricities and inclinations initially similar to those of their progenitors. 
This is in contrast to ambient nebular hydrogen which presumably traces circular heliocentric orbits. Thus the debris cloud will be moving relative to the surrounding hydrogen, at a speed equal to the cloud's non-circular heliocentric velocity. We adopt a fiducial speed for this ``nebular headwind'' 
of $u_{\rm hw} = 3$ km/s (equivalently $\sqrt{e^2+i^2} \simeq 0.15$).} The nebular headwind will distort the initially overpressured and expanding vapor cloud into a shape resembling a cometary coma in the solar wind. Our calculations below will be restricted to the debris expanding directly into the headwind (the `head' of the `comet').

The debris cloud is assumed to be dominated by vapor and to contain in lesser proportions solid ejecta of various sizes, among which are chondrule precursors. The vapor is assumed to be sufficiently decompressed that it behaves as an ideal gas; we do not model the prior non-ideal phase of the explosion \citep[cf.][]{johnson_melosh_2015, stewart_lpsc6,davies_etal_2020}. \rev{Our fiducial input parameter values 
include the vapor cloud's initial radius, temperature, and mass density $(R_0,\, T_0,\, \rho_0) = (200\,\mathrm{km},\, 4000\,\mathrm{K},\, 10^{-5}\,\gcc$); alternate parameters are explored in section \ref{subsec:param_space_cooling}.} Our analysis is restricted to zeroth-order properties of the cloud, e.g. its mean temperature and pressure. It may be that our assumption of a vapor-dominated plume is unrealistic; for a melt-dominated plume, see \cite{dullemond_etal_2014}.

We take the local density of the background hydrogen nebula to be $\rhon = 10^{-11}$ g/cm$^3$, a factor of ten lower than that of the minimum-mass solar nebula at 3 au,
and reflecting perhaps a more evolved disk. For a local nebular temperature of $T_{\rm neb} = 75$ K (unheated
by planetesimal bow shocks) and a mean molecular weight of $\mu_{\rm neb} = 2.4$ for
solar composition gas, the nebular sound speed is $c_{\rm neb} = 0.5$ km/s.

Sections \ref{subsec:cloud1}--\ref{subsec:backfill} describe,
in chronological order,
the dynamical and thermal evolution of the vapor cloud. 
Section \ref{subsec:chondrule1}
considers the evolution of 
 a proto-chondrule, treated
as a kind of test particle
in the cloud.

\vspace{0.1in}

\subsection{Overpressured expansion and adiabatic cooling}
\label{subsec:cloud1}

At the start of our calculation (time $t = 0$), the cloud pressure greatly exceeds the nebular pressure, and the nebular mass displaced by the cloud is small compared to the cloud mass; thus the cloud expands nearly freely. We assume the expansion velocity $u$ at this time is at its peak value, $u = \max u = u_0$, and is such that the cloud's
bulk kinetic energy per unit mass,
$(1/2)u_0^2$, is comparable to its 
internal energy 
$kT_0/[(\gamma-1)\mu m_{\rm p}]$ \citep{zr}: 
\begin{align}
    u_0 
    &\sim \left(\frac{2k T_0}{(\gamma-1)\mu m_{\rm p}}\right)^{1/2} \nonumber \\ 
    &\sim 2.6\,\mathrm{km}\,\mathrm{s}^{-1}\left(\frac{T_0}{4000\,\,{\rm K}}\right)^{1/2}
    \label{eqn:cs0}
\end{align}
where $k$ is Boltzmann's constant, $m_{\rm p}$ is the proton mass, and $\mu = 30$
and $\gamma = 4/3$  (\citealt{melosh_1989}) are
the mean molecular weight
and adiabatic index, respectively, 
of hot silicate vapor 
(see e.g. figure 4 of 
\citealt{fegley_schaefer_2012}). The vapor mostly comprises Na and O atoms, and SiO and O$_2$ molecules whose ro-vibrational degrees of freedom are excited. In reality $u_0$ may be somewhat higher
than given by (\ref{eqn:cs0}) because the expansion
speed at $t=0$ respects how much thermal energy is present during the cloud's inception at $t < 0$, when the
temperature may exceed $T_0$. We neglect this order-unity correction, which is also complicated by non-ideal-gas effects.

\rev{The cloud's initial expansion is supersonic with respect to the surrounding nebula, 
$u_0/c_{\rm neb} = 5$. A forward shock propagates outward, sweeping nebular hydrogen into a shell whose mass grows at a rate
\begin{equation}
    \frac{dM_{\rm neb}}{dt} = 4\pi R^2(t) \rhon [u(t) + u_{\rm hw}]
    \label{eqn:mdot}
\end{equation}
where $R(t)$ is the cloud's radius and $u(t) = \dot{R}$. In writing (\ref{eqn:mdot})
we are considering the cloud's expansion directly
against the headwind (so $u(t)$ and $u_{\rm hw}$ add), and idealizing the flow as
spherically symmetric.
The nebular gas carries an opposing momentum density $\rhon u_{\rm hw}$ and forces the cloud to decelerate according to
\begin{equation}
    \frac{d}{dt} \left\{ \left[ M_{\rm neb}(t) + M_0 \right] u(t) \right\} = -4\pi R^2 \rhon u_{\rm hw} [u(t) + u_{\rm hw}],
    \label{eqn:pdot}
\end{equation}
where $M_0 = (4\pi/3)R_0^3\rho_0$ is the conserved mass of silicate material in the cloud. We numerically integrate equations (\ref{eqn:mdot}) and (\ref{eqn:pdot}) for $R(t)$, from which the mean density of silicates 
\begin{equation}
    \rho(t) = \rho_0\left(\frac{R}{R_0}\right)^{-3}
    \label{eqn:rhot}
\end{equation} 
follows.
}

Initially, as the cloud expands, its mean temperature and pressure drop adiabatically from their starting
values $T_0$ and $P_0 = \rho_0 k T_0 / (\mu m_{\rm p})$:
\begin{align}
\frac{T}{T_0} &= \left(\frac{\rho}{\rho_0}\right)^{\gamma - 1} \\
\frac{P}{P_0} &= \left(\frac{\rho}{\rho_0}\right)^{\gamma} 
\label{eqn:adiabat}
\end{align}
with $\gamma = 4/3$ as stated earlier. 
These relations describe the silicate
cloud material, not the piled-up
nebular mass. When the vapor cools
enough to condense, $T$ and $P$ deviate
from these power-law adiabats, as will be
described in section \ref{subsec:condense}.

\rev{In Fig. \ref{fig:background}, panel a shows $R(t)$ (dashed black curve) and $\rho(t)$ (solid black curve), panel b shows $T(t)$ (solid blue curve), and panel d shows $P(t)$ (solid black curve). For $t \lesssim 4 \times 10^3$ s, the cloud expands at near-constant velocity $u \simeq u_0$ so that $R \propto t$ and $\rho \propto t^{-3}$. Around $t \sim 4 \times 10^3$ s, the expansion slows from the swept-up nebular gas. We end our calculations when $u=0$, at which point $t = t_{\rm stall} \sim 8 \times 10^3$ s, $R = R_{\rm stall} \sim 10^4$ km, and the silicate cloud density $\rho = \rho_{\rm stall}$ is within a factor of 5 of $\rho_{\rm neb}$. Section \ref{subsec:backfill} discusses what happens after $t_{\rm stall}$.
}

\subsection{Condensation}
\label{subsec:condense}

The silicate vapor departs from the original adiabat at $t \sim 40$ s, when it cools sufficiently that it starts to re-condense into liquid/solid droplets (hereafter ``dust''---not to be confused with chondrules, which will not be discussed until section \ref{subsec:chondrule1}). Condensation occurs when the saturation vapor pressure $P_{\rm sat}$ given by
\begin{equation}
\log_{10}\left(\frac{P_{\rm sat}}{\rm bars}\right) = -30.6757 - \frac{8228.146\,\,{\rm K}}{T}+\, 9.3974\log_{10}\left(\frac{T}{{\rm K}}\right)
    \label{eqn:psat}
\end{equation}
equals the cloud pressure $P$. We use here the vapor pressure for molten bulk silicate Earth (``BSE''), which has a chemical composition similar to that of olivine-rich chondrites \citep{fegley_schaefer_2012}. Once saturated, the cloud remains saturated, with the gas pressure $P$ equal to $P_{\rm sat}(T)$ \citep{zr, melosh_1989}.\footnote{The cloud may super-cool along the original adiabat before equilibrium condensation sets in. For our cloud whose expansion speed does not greatly exceed the thermal speed of its constituent gas molecules, and which may be full of impurities and condensation sites (i.e., liquid/solid ejecta from the original collision), this departure from equilibrium should be brief.}

Condensation releases latent heat, which keeps the gas warmer than it would be along the original adiabat. The temperature of the dust-gas mixture evolves from adiabatic cooling, release of latent heat, and radiation emitted by dust:
\begin{align}
    \frac{4\pi \rho R^3}{3}\biggl(\left[C_{\rm gas}(1-x) + C_{\rm solid}x\right]dT - \frac{k}{\mu m_{\rm p}}T \left(1-x\right)\frac{d\rho}{\rho}  \nonumber  \\  - \left[L_{\rm vap} - (C_{\rm solid} - C_{\rm gas})T \right]dx \biggr)   = - \frac{4\pi R^2 \sigma_{\rm SB}T^4}{\tau}dt \,.
    \label{eqn:condensation_energy}
\end{align}
The left-hand side of (\ref{eqn:condensation_energy}) is taken from \citet[][chapter 8]{zr}, where $x \equiv \rho_{\rm dust}/\rho$ is the mass fraction in condensates, $C_{\rm gas} = 6.3 \times 10^6$ erg/(g K) is the gas specific heat at constant volume \citep{melosh_1989}, $C_{\rm solid} = 10^7$ erg/(g K) is the specific heat of dust, and $L_{\rm vap} = 3 \times 10^{10}$ erg/g is the heat of silicate vaporization \citep[measured for pure forsterite;][]{nagahara_etal_1994}. Whereas \citet{zr} have no right-hand side term because they assume strict energy conservation,
we account in our right-hand side for thermal photons that diffuse out of the assumed optically thick cloud, with $\sigma_{\rm SB}$ equal to the
Stefan-Boltzmann constant and
\begin{align}
\tau &=  \frac{3\rho R}{4\rhos s_{\rm dust}}x 
\label{eqn:tau}
\end{align}
equal to the cloud radial optical
depth, assuming that vapor condenses into dust grains of typical radius $s_{\rm dust}$ and internal density $\rho_{\rm solid} = 3\,\gcc$. Each dust grain is assumed to present a geometric cross section to photons. Our radiative loss term in (\ref{eqn:condensation_energy}) presumes that the cloud is in radiative equilibrium,
with the bulk of the dust at temperature $T$ heating photospheric dust near the
cloud outer boundary 
to a temperature of 
$\sim$$T/\tau^{1/4}$.
We take $s_{\rm dust} = 1\,\mu$m, comparable in
size to vapor condensates 
in other settings,
including terrestrial experiments of condensing silicate and metal vapor  
\citep[][page 70]{melosh_1989},
silicate
clouds in exoplanet
atmospheres \citep[e.g.][]{gao_etal_2020_nature}, and vapor plumes of meteors impacting the Earth at $\lesssim 15$ km/s
\citep[][their figure 13]{johnson_melosh_2012}. We
test different values of $s_{\rm dust}$ in section \ref{subsec:param_space_cooling}.

Equation (\ref{eqn:condensation_energy})
is solved numerically for $T(t)$ and $x(t)$, 
in conjunction with Eqs. (\ref{eqn:mdot})--(\ref{eqn:rhot}) for $R(t)$ and $\rho(t)$, and the condition
$P = (1-x)\rho k T/(\mu m_{\rm p}) = P_{\rm sat}(T)$. 
\reva{Fig. \ref{fig:background} shows how, at $t \sim 40$ s, our fiducial cloud cools to $T \approx 2600$ K, at which point $P_{\rm sat}(T)$ crosses and subsequently locks to $P$ (panel d). From here on out, dust condenses out of vapor and $x$ grows from zero. Note the system temperature $T \approx 2600$ K at this time exceeds the value of $T \approx 1500$ K often taken to signal condensation. The latter is valid at low pressures, not the high pressures that characterize the impact plume. Our approach of comparing $P$ to $P_{\rm sat}$ to decide when liquid droplets can condense out of vapor is more general than using a fixed temperature condition.
}

Over the next few minutes an order-unity fraction of
the vapor condenses, rendering the
cloud optically thick (panel c)
and causing its temperature
to drop more slowly with time
than it did along the original
adiabat because of latent
heat release (panel b).
We have verified that
the cloud is in radiative
equilibrium 
insofar as the photon
diffusion time $\tau R/c$, 
where $c$ is the 
speed of light, is either 
comparable to or less than the 
elapsed time $t$.
We have also checked
that gas and dust conduct
heat to one another so 
efficiently (via gas-dust collisions) that both species are
at very nearly the same
temperature at any given time.

\subsection{Radiative losses and nebular backfilling} 
\label{subsec:backfill}

Eventually, as the cloud expands and becomes less optically thick, an order-unity fraction of the cloud's thermal energy is lost to radiation. From Fig. \ref{fig:background}b, we see that radiative losses become significant at $t \sim 3000$ s. 
After this time, $T$ declines faster than before, scaling somewhere between $T \propto  s_{\rm dust}^{-1/3}t^{-5/3}$ and $T \propto s_{\rm dust}^{-1/3}t^{-1/3}$, as can be seen from equation (\ref{eqn:condensation_energy}) by ignoring the volume expansion and latent heat terms, taking a constant $x \sim 1$, and using either $R \propto t$ (free expansion) or $R \propto t^0$ (stalled expansion). The faster temperature decline causes residual silicate vapor to condense and the gas density to plummet (Fig. \ref{fig:background}a).

Radiation losses and condensation thus cause the cloud's internal pressure to fall dramatically below the external nebular pressure (Fig. \ref{fig:background}d). The cloud's momentum allows it to continue expanding against this adverse pressure gradient until $t_{\rm stall} \sim 8 \times 10^3$ s, when it comes to a stop and our calculation formally ends. After this point, we expect the headwind of nebular hydrogen to backfill the nearly pressure-free cavity, sweeping past whatever large fragments remain from the collision and carrying away chondrules and small condensates at speed $u_{\rm hw}$ (for more on this, see Section \ref{sec:agglom}). \cite{stewart_lpsc6} miss this unidirectional headwind because their collision target was assumed unrealistically to be at rest relative to the nebular gas. The plume originating from their target thus undergoes a nearly spherical collapse, when it should be swept away by the headwind.


\subsection{Chondrule thermal histories}
\label{subsec:chondrule1}

Having described the evolution of the silicate cloud, we now examine how chondrules may be created within the cloud. We assume at $t=0$ that the cloud, of size $R_0$, contains chondrule precursors, modeled as spheres of internal density $\rhos$ and radius $\sch$. \reva{These precursors were either condensed from the gas phase --- as may have been the case for CB chondrules --- or were ejected from the collision in solid or partially molten form and may thus contain relict grains.}

It is not obvious that liquid/solid particles can survive without vaporizing if embedded in the cloud at $t < 0$, when $R < R_0$, $T > T_0$, and $\rho > \rho_0$. However, the collisional destruction of planetesimals is not instantaneous. It unfolds over the finite interval of time it takes the impactors to finish smashing through each other. For at least this initial smash-through phase, which lasts $\mathcal{O}(1\,{\rm min})$ for $\sim$100 km-sized planetesimals, debris should be continuously generated and released into the vapor cloud. While debris that is released when the impactors first make contact may not last to $t=0$, liquid/solid particles that are released towards the end of the smash-through phase, just before $t=0$, may survive.

\reva{As we do not model the complicated dynamics of the collision \citep[cf.][]{stewart_lpsc7}, we cannot determine how many 
chondrule precursors are created by gas-phase condensation or are ejected directly from the planetesimals and survive vaporization.} We proceed by assuming that such particles exist and are entrained by the cloud, and that their combined mass is less than the cloud mass ($=4\pi R_0^3 \rho_0/3$), so that we may neglect how the particles affect the cloud's expansion. Order-of-magnitude estimates suggest these particles have chondrule-like sizes. A liquid droplet moving through vapor must be small enough that the surface tension force holding it together, $\sim$$2\pi \sigma \sch$, exceeds the disruptive force of aerodynamic drag, $\sim$$\rho_0 \sch^2 u_0^2/2$ (\citealt{melosh_1989}). For our fiducial parameters, this limiting size is $\sch \sim 4\pi \sigma/(\rho_0 u_0^2) \sim \mathcal{O}(0.1\,\rm mm)$, for a surface tension $\sigma \sim 350$ dyne/cm appropriate to molten rock. Using similar arguments, \cite{melosh_vickery_1991} and \cite{johnson_melosh_2014} found that molten ejecta are shredded into millimeter-scale droplets. Particles of this size are readily entrained in the vapor cloud; right at the outset of the cloud's expansion, the aerodynamic stopping times $t_{\rm stop} \sim \rho_{\rm solid} \sch^3 u/F_{\rm drag}$ of mm-sized particles are orders of magnitude shorter than the cloud dynamical time $R/u$, 
for a drag force $F_{\rm drag} \sim \rho u^2 \sch^2$.

We now describe the thermal histories of these entrained particles.

\subsubsection{Chondrule thermal history, fiducial asteroid belt model}
Proto-chondrules exchange heat with their environment by gas conduction and radiation. Their temperature $T_{\rm c}$ evolves as:
\begin{align}
\frac{4\pi}{3} \sch^3 \rhos C_{\rm solid} \frac{d\Tc}{dt} &= 
    n_{\rm gas} \pi \sch^2 u_{\rm th} k(T - \Tc)  +4\pi \sch^2 \sigma_{\rm SB} (T^4  - \Tc^4) 
\label{eqn:edot_ch}
\end{align}
where $n_{\rm gas} = (1-x)\rho/(\mu m_{\rm p})$ is the number density of gas molecules, $u_{\rm th} = \sqrt{8kT/(\pi \mu m_{\rm p})}$ is the gas mean thermal speed, and each collision between a gas molecule and a chondrule is assumed to transfer an energy $k(T - \Tc)$. The heating term proportional to $\sigma_{\rm SB} T^4$ is due to the background radiation field emitted by optically thick dust (after it condenses). What equation (\ref{eqn:edot_ch}) omits is drag-heating by the silicate vapor, but this effect lasts only briefly, for the fraction of a second
it takes the chondrule to come up
to speed with the cloud, and even
then adds only marginally to conductive
heating.

\rev{We set the proto-chondrule's initial
temperature $T_{\rm c}(0) = 10^3$ K,
a value within the wide range of temperatures
to which solids are heated upon impact
(depending principally on distance from the impact site; see figure 2 of \citealt{johnson_melosh_2014}).
This initial temperature is quickly forgotten as the proto-chondrule comes into thermal and dynamical equilibrium with the cloud.
Equation (\ref{eqn:edot_ch}) is solved for $T_{\rm c}(t)$ using a fiducial chondrule radius $s_{\rm c} = 0.3$ mm, with the background variables $n_{\rm gas}(t)$ and $T(t)$ calculated separately as described in previous sections. }

\begin{figure}
\vspace{-1.5cm}
\includegraphics[width=0.95\columnwidth]{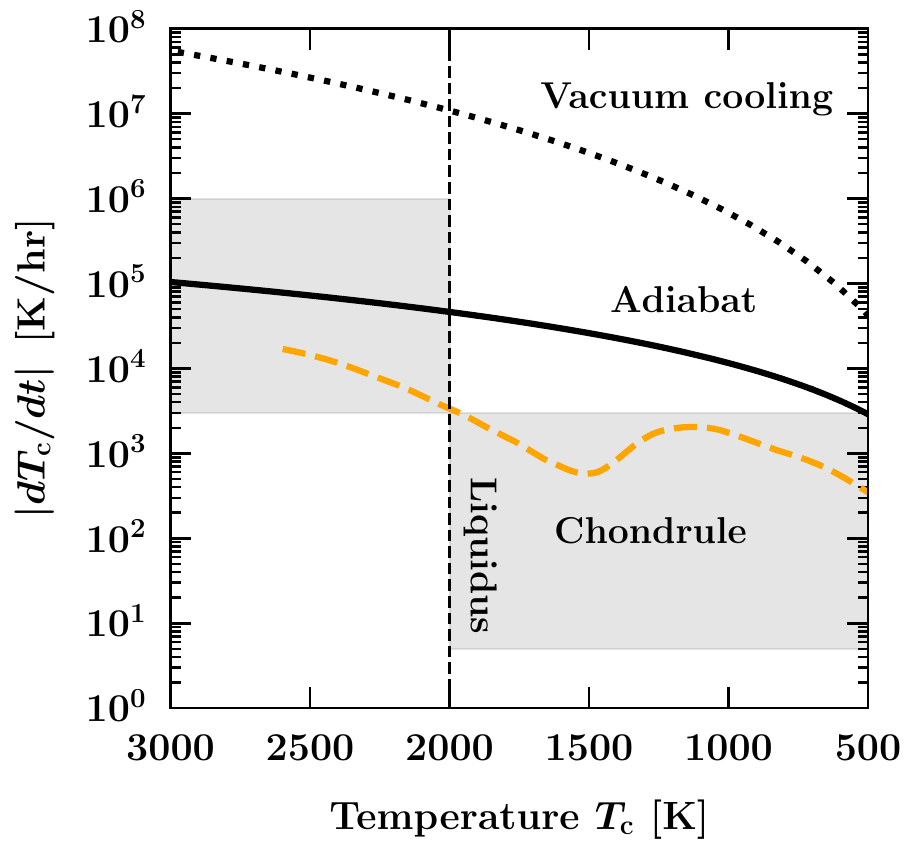}
\caption{ Chondrule cooling rate vs. temperature (dashed orange curve), shown from the point the cloud saturates (in this regime radiation by newly condensed dust enforces $T_{\rm c} = T$, and $T_{\rm c}$ falls monotonically). From $T_{\rm c} = 2600$ K down to $T_{\rm c} = 1500$ K, cooling is mostly driven by the cloud's expansion, with cooling rates lower than given by the original
adiabat (solid curve) which
does not account for latent
heat released by condensation. Also shown for comparison is the blackbody cooling rate
in vacuum ($dT_{\rm c}/dt = 4\pi s_{\rm c}^2\sigma_{\rm SB}T_{\rm c}^4/(4\pi \rho_{\rm solid} s_{\rm c}^3 C_{\rm solid}/3)$; dotted curve).  For $T_{\rm c} \lesssim 1500$ K, radiative losses from the cloud as a whole become important and the cooling rate increases. 
\eugene{The shaded regions highlight super and sub-liquidus cooling rates inferred from laboratory experiments. These constraints are taken from \protect\cite{desch_connolly_2002}, except for the poorly constrained upper bound to the super-liquidus cooling rate which we arbitrarily set at $10^6$ K/hr.} 
}
  \label{fig:Tdot}
\end{figure}

Fig. \ref{fig:background}b shows $T_{\rm c}(t)$. Initially, conductive heating is balanced by radiative cooling into the optically thin cloud (dust has not yet condensed), and the chondrule is at a temperature of  $T_{\rm c} \sim 2000$ K. The saturation vapor pressure
of the chondrule, $P_{\rm sat}(T_{\rm c})$ (equation \ref{eqn:psat}), sits more than two orders of magnitude below the ambient gas pressure $P$ (Fig. \ref{fig:background}d), safeguarding the chondrule against vaporization (for a study of the time-dependent kinetics of vaporization and volatile retention, see \citealt{dullemond_etal_2016}). Over the next $\sim$30 s,
the chondrule remains colder than, but cools in lockstep with, the adiabatically expanding background gas, falling to $T_{\rm c} \sim 1500$ K.
At $t \sim 40$ s, the chondrule heats 
back up to $\sim$2600 K
when dust condenses and
renders the entire cloud optically
thick --- the chondrule is
literally ``flash-heated'' by
the radiation emitted by newly
condensed dust \reva{(as discussed 
in section \ref{subsec:condense},
condensation is possible for $T$ 
as high as $2600$ K because the
high cloud pressures cross
the saturation vapor pressure).
}
The chondrule is now trapped
in this radiation bath and
$T_{\rm c} \simeq T$. Over
the course of $\sim$5 minutes,
the temperature falls from $\sim$$2600$ K
through the
liquidus of $\sim$2000 K (below which melt and solid co-exist); it then passes 
through the solidus,
here estimated to be $\sim$1500 K (below which the particle is entirely solid),
after $\sim$30 minutes.
Cooling rates $|dT_{\rm c}/dt|$
are plotted versus $t$ in Fig. \ref{fig:background}b
and versus $T_{\rm c}$ in
Fig. \ref{fig:Tdot}.
Above the liquidus, $|dT_{\rm c}/dt| \sim 3000$--20000 K/hr, while 
below the liquidus $|dT_{\rm c}/dt| \sim 300$--3000 K/hr. These cooling rates appear compatible with 
empirically determined chondrule cooling rates \citep[e.g.][]{desch_connolly_2002, connolly_jones_review},
shown as grey regions in 
Fig. \ref{fig:Tdot}. \reva{They also overlap with the cooling rates determined empirically for CB chondrules by \cite{hewins_etal_2018}.} Fig. \ref{fig:Tdot} shows that adiabatic cooling alone 
predicts cooling rates that exceed sub-liquidus experimental rates by at least an order of magnitude; cooling buffered
by dust condensation is essential
to reproducing chondrule cooling rates.


\subsubsection{Cooling rate variations over parameter space}
\label{subsec:param_space_cooling}

\begin{figure}
\vspace{-2.0cm}
\includegraphics[width=\columnwidth]{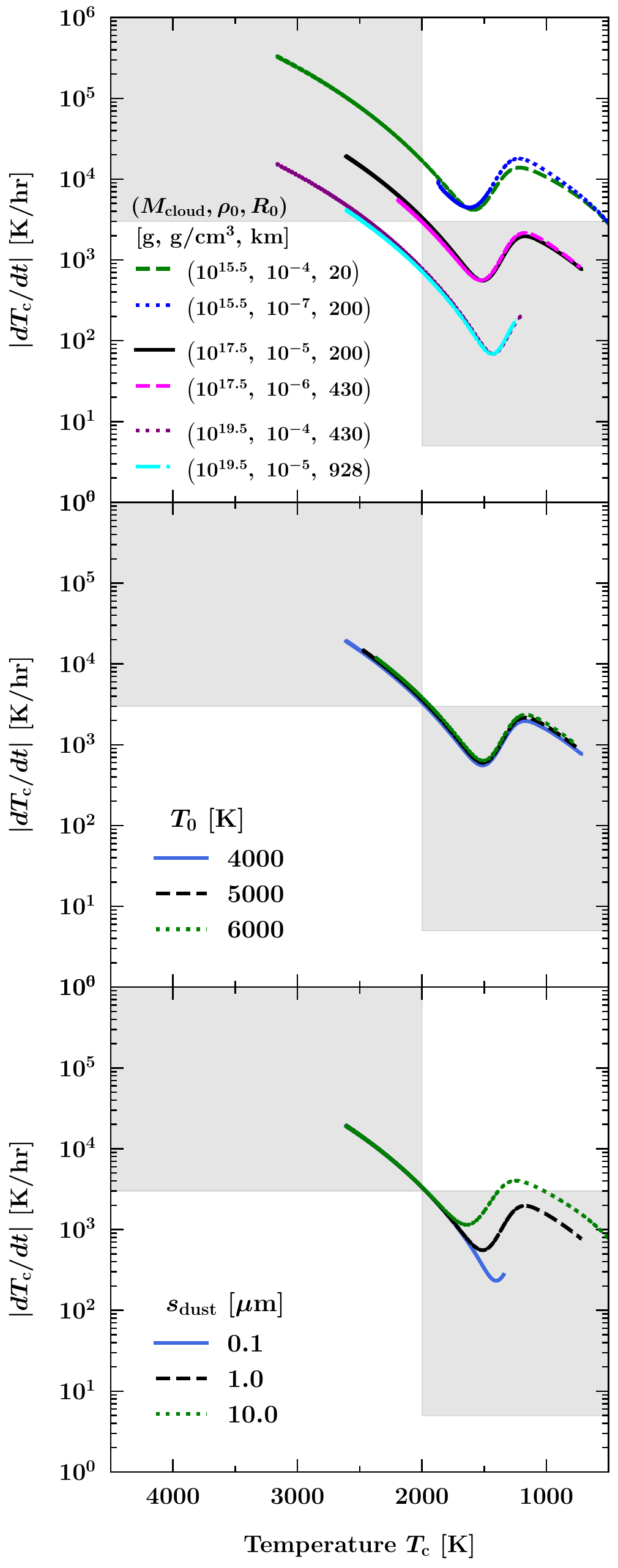}
\vspace{-0.6cm}
\caption{ \rev{\textit{Top:} Chondrule cooling rates vs. temperature (as in Fig. \ref{fig:Tdot}) for various initial cloud densities $\rho_0$ and radii $R_0$. Cooling rates scale with the total cloud mass, $dT_{\rm c}/dt \propto M_{\rm cloud}^{-1/3}$ (equation \ref{eqn:Tdot_Mcloud}), where $M_{\rm cloud} \sim \rho_0 R_0^3$. Cooling rates satisfy experimental constraints (shaded regions) for $M_{\rm cloud} \gtrsim 10^{17}$ g. \textit{Middle:} Cooling rates for different initial cloud temperatures $T_0$. The value of $T_0$ has little effect on cooling rates because cooling is governed by the equilibrium condition $P = P_{\rm sat} (T)$
which is independent of initial conditions. \textit{Bottom:}  Cooling rates for different assumed radii $s_{\rm dust}$ of condensed dust grains. The dust size does not matter at higher temperatures (early times) when the cloud cools by expanding and loses negligible energy to radiation. Larger dust grains render the cloud less optically thick and hasten the onset of radiative
losses, which increase cooling rates. }}
\label{fig:Tdot_T_params}
\end{figure}

We explore how chondrule cooling rates
change with initial cloud properties $\rho_0$, $R_0$, and $T_0$, 
as well as the 
sizes of condensed
dust grains $s_{\rm dust}$ and the 
background
nebular density $\rho_{\rm neb}$. 
Unless otherwise indicated, 
we vary one parameter at a time while holding others fixed at their fiducial values 
($\rho_0, R_0, T_0, \rho_{\rm neb}, s_{\rm dust}) = (10^{-5} \,\gcc, 200\,\mathrm{km}, 4000\,{\rm K}, 10^{-11}\, \gcc, 1 \,\mu{\rm m})$.

Fig. \ref{fig:Tdot_T_params} plots chondrule cooling rates starting from when vapor saturates, dust forms, and the chondrule temperature $T_{\rm c}$, radiatively locked to the cloud temperature $T$, falls monotonically (see Figure  \ref{fig:Tdot}).
Fig. \ref{fig:Tdot_T_params}a varies the cloud's initial radius $R_0$ and initial density $\rho_0$, and demonstrates that what matters for the
cooling rate $dT_{\rm c}/dt$ at a given temperature $T$ 
is the product $\rho_0 R_0^3$, i.e., the total cloud mass $M_{\rm cloud} = 4\pi \rho_0 R_0^3/3$. 
This dependence follows from
\begin{equation}
\frac{dT_{\rm c}}{dt} = \frac{dT}{dt} = \frac{dT}{d\rho}  \frac{d\rho}{dR} \frac{dR}{dt} \,.
\label{eqn:Tdot_Mcloud}
\end{equation}
The factor $dT/d\rho$ depends only the equilibrium thermodynamics of adiabatic cooling and condensation, which specifies $\rho(T)$ (equation \ref{eqn:condensation_energy}); it does not depend on the initial conditions $\rho_0$ or $R_0$. The second factor $d\rho/dR \propto \rho/R$ (equation \ref{eqn:rhot}), which for given $\rho(T)$ scales as $1/R \propto M_{\rm cloud}^{-1/3}$. 
\rev{The final factor $dR/dt$ is nearly constant ($=u_0$) at early times when the cloud is freely expanding, and depends only on $\rho(T)$ and $u_{\rm hw}$ in the final stages of the expansion just before the cloud stalls.} Putting it all together, we see that $dT_{\rm c}/dt$ scales as $M_{\rm cloud}^{-1/3}$ at all times;
a more massive cloud cools
more slowly because to reach
a given $\rho(T)$ it needs
to expand to a larger radius $R$,
when its dynamical time
$R/\dot{R}$ is longer.

\reva{According to Fig. \ref{fig:Tdot_T_params}a, cloud masses $M_{\rm cloud} \sim 10^{17}$--$10^{20}$ g yield chondrule sub-liquidus cooling rates consistent with the fastest rates inferred from petrologic experiments, in particular the $\sim$100--1000 K/hr rates inferred for skeletal, non-porphyritic CB chondrules \citep{hewins_etal_2018}.
The impact cloud can only be less massive than the colliding asteroids from which it derives; 
$M_{\rm cloud} \sim 10^{17}$--$10^{20}$ g imposes hard lower limits on the combined radius of the colliding planetesimals of 
$R_{\rm pl} \gtrsim \left[3M_{\rm cloud}/(4\pi \rhos)\right]^{1/3} \sim 2$--20 km. If we assume a vapor production efficiency of 1\% by mass (cf. \citealt{johnson_melosh_2015,wakita_etal_2017, wakita_etal_2021}; these studies technically track melt and not vapor), the colliding planetesimals would be 10--100 km in radius, similar in size to the asteroids that contain most of the main belt mass today, and perhaps also in the past \citep{morbidelli_etal_2009}.} 

The value of $T_0$ does not much  affect the cooling rate  post-saturation (Fig. \ref{fig:Tdot_T_params}b), when $P=P_{\rm sat} (T)$ and the cloud thermodynamics evolves in an equilibrium
fashion with initial conditions largely forgotten. What small differences can be seen in  Fig. \ref{fig:Tdot_T_params}b for $dT_{\rm c}/dt$ arise from variations in the free-expansion cloud velocity $u_0 \propto T_0^{1/2}$ (equation \ref{eqn:cs0}).

Fig. \ref{fig:Tdot_T_params}c
varies $s_{\rm dust}$, the assumed radii of dust grains that condense out of the silicate vapor.
At the earliest times, when cooling is driven by expansion and not by radiation, $s_{\rm dust}$ is irrelevant.
Radiation becomes important sooner, and cools the cloud faster,
for larger $s_{\rm dust}$ which makes the cloud less optically thick
(equation \ref{eqn:tau}). The radiation-dominated cooling rate $dT_{\rm c}/dt = dT/dt$ at a given $T$ (not $t$) scales between $s_{\rm dust}^{1/5}$ during free expansion and $s_{\rm dust}^1$ as the cloud stalls. This can be seen from equation (\ref{eqn:condensation_energy}) in combination with the scalings $T \propto s_{\rm dust}^{-1/3} t^{-5/3}$ and $T \propto s_{\rm dust}^{-1/3} t^{-1/3}$ as derived in section \ref{subsec:backfill}, with $t$ substituted in favour of $T$.

\subsubsection{Collisions in the proto-Kuiper belt} 
\label{subsec:kuiper}

\begin{figure*}
\includegraphics[width=\textwidth]{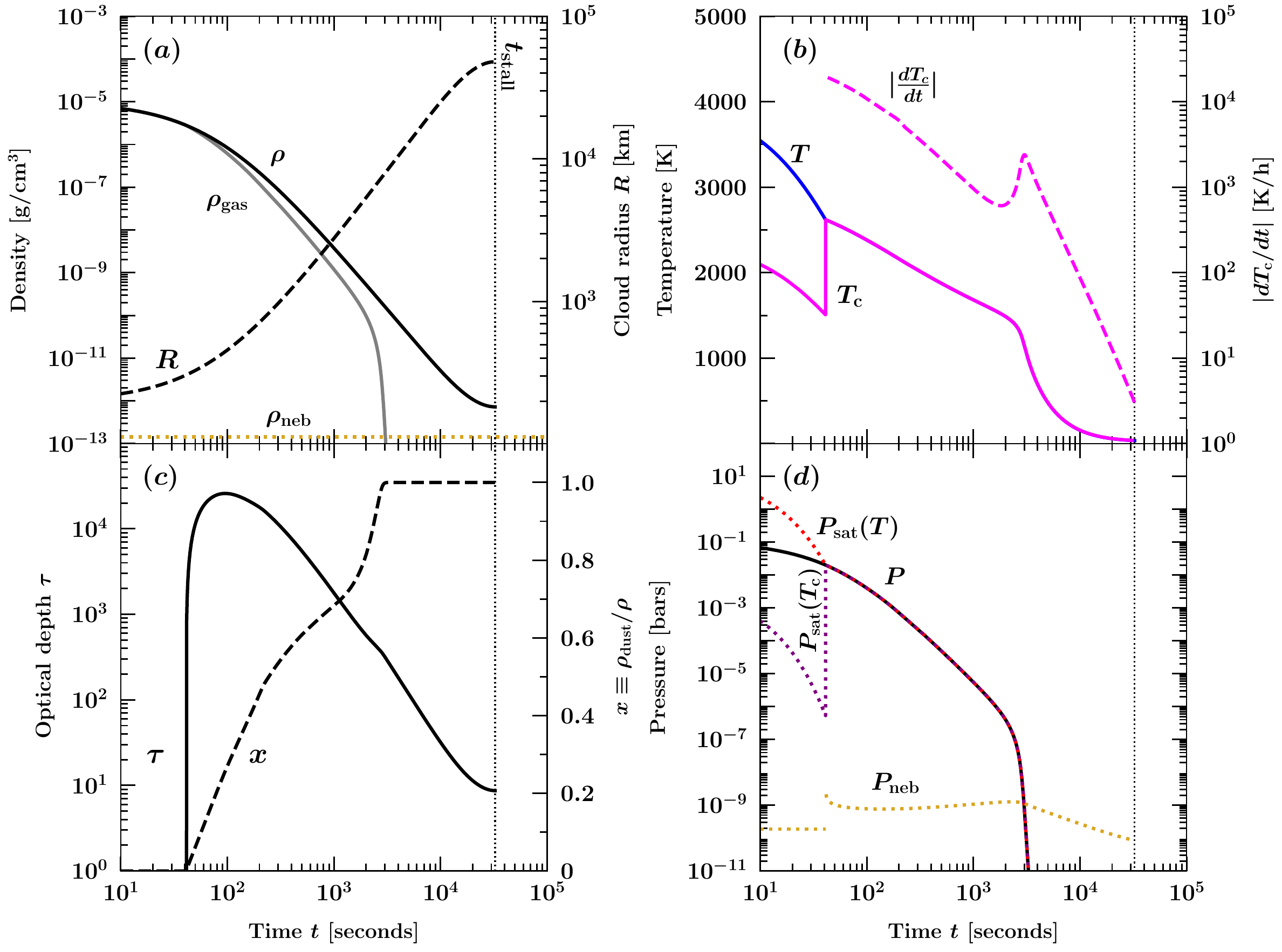}
\caption{Same as Fig. \ref{fig:background}, but for a vapor cloud created at $a = 15$ au in a nebula of density $\rhon = 1.1 \times 10^{-13}\,\gcc$, conditions intended to model those of the proto-Kuiper belt where comet Wild-2 may have formed. The change in scenery does not much affect the thermal evolution of either the silicate cloud or the chondrule at temperatures $T \gtrsim 1000$ K and times $t \lesssim 5 \times 10^3$ s.
The main difference in cloud evolution
between asteroid belt and Kuiper belt distances is at late times. \rev{Because the nebular density $\rhon$
is lower at 15 au than at 3 au, the nebula is less effective at slowing down the cloud, which does not stall until $t_{\rm stall} = 3.5 \times 10^4$ seconds, a factor of 4 later than the stalling time at 3 au. }}
  \label{fig:kuiper_4pan}
\end{figure*}

\rev{The {\it Stardust} spacecraft 
discovered that the short-period comet 81P/Wild-2 
contains chondrule-like particles 
\citep[e.g.][]{nakamura_etal_2008}.
By ``chondrule-like'' we mean igneous
particles having mineralogies and textures,
and by extension thermal histories, 
similar to those of asteroidal chondrules.
The particles collected have sizes up to
$\sim$$10 \,\mu$m. Sampling larger particles would have required the {\it Stardust} spacecraft to approach closer to the comet than was considered safe.}

To assess whether high-velocity collisions can explain the heating experienced by these cometary particles, we re-scale
our model to heliocentric distances possibly appropriate to Wild-2's formation. 
Short-period comets like Wild-2 originate as Kuiper belt objects (KBOs), predominantly of the ``scattered'' variety having relatively large orbital eccentricities and inclinations \citep[e.g.][]{nesvorny_etal_2017}. Dynamically hot KBOs are thought to reflect a period of upheaval when the orbits of the giant planets (and perhaps those of planets no longer present) underwent large-scale changes driven by gravitational scatterings with remnant planetesimals \citep[e.g.][]{fernandez_ip_1984, malhotra_1993, tsiganis_etal_2005, gomes_etal_2005, ford_chiang_2007, levison_etal_2011, dawson_murray-clay_2012}.
During this time, Neptune and proto-KBOs were propelled
outward from $\sim$10--20 au 
to 30 au and beyond.
Accordingly, we re-stage our calculations for $a = 15$
au, \rev{near where proto-KBOs may have originated. We adopt a nebular density $\rhon = 1.1 \times 10^{-13}\,\gcc$, a factor of 90 lower than our fiducial value at 3 au, as follows from the scaling law $\rhon \propto a^{-39/14}$ derived for the solar nebula \citep[e.g.][]{chiang_youdin_2010}.} 
\rev{For simplicity we keep all other
model parameters fixed at their fiducial
values; this assumes the lower
heliocentric velocity at 15 au vs. 3 au
is compensated by higher orbital eccentricities and/or inclinations, to keep
the relative collision velocity between
planetesimals, and by extension the nebular headwind velocity $u_{\rm hw}$, unchanged.
Note also that we keep our fiducial
chondrule size at $\sch = 0.3$ mm,
as this is still presumably the characteristic
melt droplet size set by the balance between surface
tension and ram pressure disruption by the expanding vapor (section \ref{subsec:chondrule1}).}
 
\begin{figure}
\includegraphics[width=0.95\columnwidth]{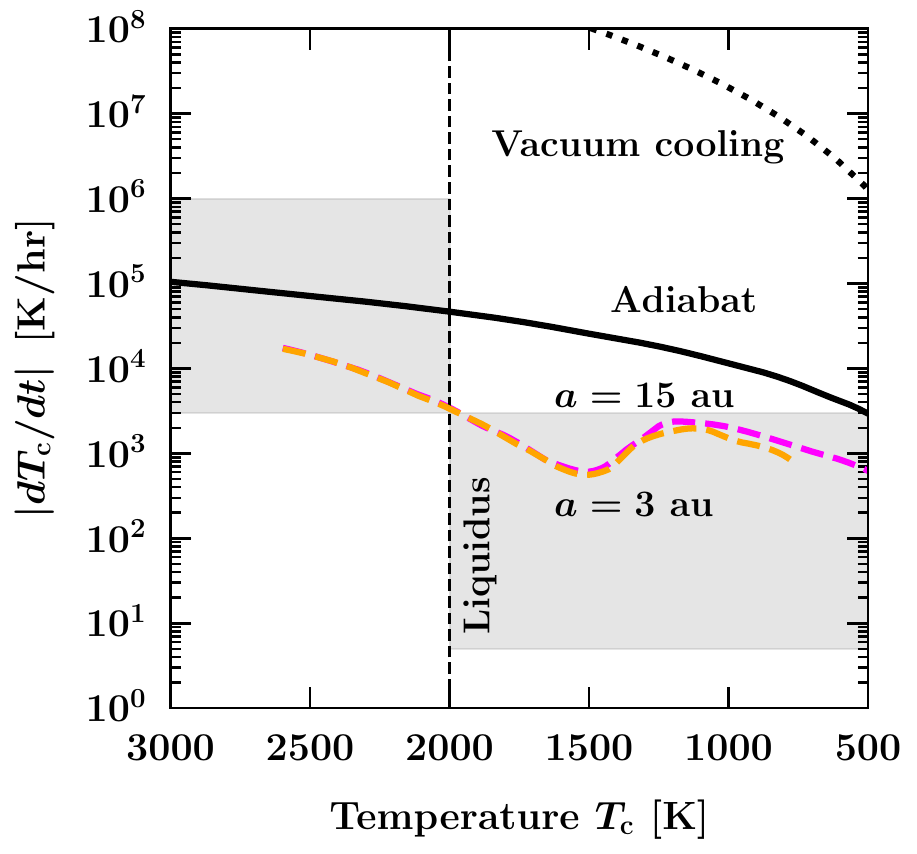}
\caption{Comparison of chondrule cooling rates at 3 au in the asteroid belt (as in Fig. \ref{fig:Tdot}) and at 15 au in the proto-Kuiper belt. Collisional vapor clouds have thermal histories that are practically
independent of heliocentric distance and might thus explain the thermal histories of chondrule-like particles discovered by \textit{Stardust}. Cooling is somewhat faster at 15 au than at 3 au because at larger distances the cloud is less impeded by a less dense nebula, and so expands faster.}
  \label{fig:Tdot_kuiper}
\end{figure}

Comparison of Fig. \ref{fig:background} with Fig. \ref{fig:kuiper_4pan} demonstrates what one might have expected: for a given set of initial cloud conditions ($\rho_0$, $T_0$, $R_0$), thermal histories
of cloud-embedded particles at 15 au are qualitatively
the same as at 3 au.
The chondrule thermal evolution
is controlled by the 
internal thermodynamics of the
vapor cloud which
are not especially sensitive 
to nebular environment, especially
when the cloud is still hot and the chondrule is passing through
the liquidus. This is further evidenced in
Fig. \ref{fig:Tdot_kuiper}.
Sub-liquidus cooling rates $dT_{\rm c}/dt$
for a given $T$
are marginally faster in the proto-Kuiper belt than in the asteroid belt because the
lower nebular density at larger heliocentric distance allows the cloud to expand freely for longer. 

While we have so far focused on generating chondrule-like thermal histories, comets like Wild-2 are replete 
with other kinds of 
equilibrated aggregates (EAs)
which were also once partially molten,
but which are smaller (0.1--1 $\mu$m), attained lower peak temperatures ($\sim$1200 K), and skew
toward faster cooling rates ($\gtrsim 500$ K/hr; 
\citealt{bradley_1994, brownlee_etal_2005, messenger_nguyen_2013}). In our model, cooling rates scale inversely with the cloud mass, $dT/dt \propto M_{\rm cloud}^{-1/3}$, with $M_{\rm cloud} \gtrsim 10^{17}$ g required to produce the cooling rates exhibited by chondrules (section \ref{subsec:param_space_cooling}). 
Lower mass clouds, in the range $M_{\rm cloud} \sim 10^{15}$--$10^{17}$ g, might have hosted the faster cooling
EAs; cloud masses and peak temperatures may have been systematically
smaller in the proto-Kuiper
belt where orbital velocities were slower, and collisions less violent, than in the asteroid belt.

\section{\rev{Agglomeration into meteorite parent bodies}}
\label{sec:agglom}
In this section we assess various avenues for agglomerating chondrules into larger bodies. We continue in the context of the aftermath of a high-speed collision \reva{between two roughly equal-mass planetesimals}, though some of our ideas may apply to other chondrule formation scenarios.

We return to our fiducial asteroid belt model and
begin our consideration of agglomeration at time
$t_{\rm stall} = 8 \times 10^3$ s,
when dust and chondrules are dispersed
across a cloud of size $R_{\rm stall} \sim 10^4$ km.  
\reva{Because one or both of the planetesimals that originally collided to produce the cloud were moving on eccentric and inclined orbits}, the debris cloud will initially trace an eccentric and inclined orbit as well. Accordingly 
the debris, including large post-collision fragments, will 
encounter a headwind of nebular hydrogen at 
speed $u_{\rm hw} \sim 3 \, {\rm km/s} \, (\sqrt{e^2+i^2}/0.15) \, (3 \, {\rm AU} / a)^{1/2}$, assuming that the nebular gas occupies circular orbits. Just after $t_{\rm stall}$,
the headwind will flood the near pressure-less
volume occupied by dust, chondrules, and post-collision fragments, filling it over a timescale $R_{\rm stall}/u_{\rm hw} \sim 3 \times 10^3$ s --- probably somewhat 
longer, since the inertia of the stalled debris
cloud, whose mean density $\rho_{\rm stall}$ is $\sim$5 times higher
than the nebular headwind density $\rhon$,
will slow the headwind by an order-unity factor.

Newborn chondrules become entrained in the supersonically streaming nebular headwind over a momentum stopping time 
\begin{align}
t_{\rm stop} &\sim \frac{\rhos}{\rhon}\frac{s_{\rm c}}{u_{\rm hw}}\nonumber  \\
&\sim 3 \times 10^4\,\mathrm{s}\left(\frac{\rhon}{10^{-11}\,\gcc}\right)^{-1}\left(\frac{s_{\rm c}}{0.3\,\rm mm}\right) \left(\frac{u_{\rm hw}}{3\,\rm km/s}\right)^{-1} \,.
\end{align}
Over this aerodynamic drag timescale, chondrules produced by the collision are swept into the background nebula and join its circular motion. Large post-collision fragments cannot circularize as quickly as chondrules and dust do, and will continue moving on eccentric orbits, at least initially.

\subsection{Agglomeration onto existing planetesimals}

We first ask whether newly formed chondrules can be re-accreted onto the post-collision remains of their progenitor asteroids. For concreteness, we consider a remnant solid body (possibly a collection of re-assembled fragments) of radius $R_{\rm pl} \sim 100$ km, typical of the asteroid and Kuiper belt today \citep{morbidelli_etal_2009, sheppard_trujillo_2010}.

The remnant asteroid's cross-section for accreting chondrules is geometric and not enhanced by gravitational focusing. This is because relative velocities between the asteroid and nebula-entrained chondrules are of order $u_{\rm hw}$, which is likely greater than the asteroid surface escape velocity $u_{\rm esc} \sim 100$ m/s. 
In principle, gas drag can abet
gravity to enlarge accretion cross sections (``pebble accretion''; \citealt{ormel_klahr_2010}), but our standard $u_{\rm hw} \sim 3$ km/s headwind sweeps chondrules
past the body too quickly for gravity
to be significant. \reva{One could entertain slower headwinds by colliding planetesimals moving on more circular, less inclined orbits; but for $u_{\rm hw}$ to be less than $u_{\rm esc} \sim 100$ m/s, eccentricities and inclinations would have to be $< 5 \times 10^{-3}$, which runs counter to our premise of 
high-velocity collisions in a dynamically hot belt \citep{raymond_izidoro_2017, carter_stewart_2020, raymond_nesvorny_2020}.}

Since gravitational focusing is defunct,
of the chondrules dispersed from the
collision over a scale $R_{\rm stall}$,
only a tiny fraction 
\begin{align}
    f_{\rm agglom} &\sim \left(\frac{R_{\rm pl}}{R_{\rm stall}}\right)^2 \sim 10^{-4}
    \label{eqn:fagg}
\end{align}
can be re-accreted, corresponding to those chondrules 
that happen to lie directly in the path of the
remnant asteroid. \reva{Our value of $f_{\rm agglom}$ is orders of magnitude lower than the re-accretion efficiency of $\sim$$0.03$ estimated by Morris et al.~(\citeyear{morris_etal_2015}, their section 3.3). Their analysis, which assumes that chondrules are ``swept up'' by the remnant asteroid at speeds ranging from 25--500 m/s (their sweep-up velocities are equivalent to our headwind velocities), overestimates gravitational focusing. Their factor by which the accretion cross section is enhanced by gravity is evaluated using a ``random'', presumably particle-particle velocity of 1 m/s, when it should use instead the particle-asteroid relative velocity, which is at least 25 m/s according to their model. Thus their estimate for the gravitational focusing factor is internally inconsistent; it is too large by a factor of at least $25^2 = 625$. 
Correcting for this implies the depth of their re-accreted particle layer is unacceptably thin, on the order of a millimeter, not a meter as claimed.}


While we have shown that a given chondrule is unlikely to be re-accreted by the remains of its progenitor asteroids, it could, in principle, be accreted by another body elsewhere. 
\reva{Impact velocities would have to be low enough to avoid shattering the chondrule. Laboratory experiments find mm-sized particles fragment at impact speeds as low as $\sim$25 m/s (\citealt{wurm_etal_2005, teiser_etal_2009}).\footnote{Chondrule fragments are occasionally observed \citep[e.g.][]{nelson_rubin_2002} and chondrule rims and matrix may be shattered chondrules \citep{alexander_etal_1989}. But the fact remains that the overwhelming majority of chondrules are round and intact.} Such relative speeds, which require $e,i \lesssim 10^{-3}$, might be achieved for planetesimal accretors small enough that the circularizing/flattening effects of nebular gas drag overwhelm 
gravitational stirring by larger bodies. These small-body accretors would be much smaller than the original colliding pair of eccentric/inclined planetesimals. Note that even if the small-body accretors have vanishing $e,i$, there is a floor on the chondrule-accretor relative velocity of $c_{\rm neb}^2/(2u_{\rm K}) \sim 10$ m/s which arises because disc gas and its entrained chondrules rotate at sub-Keplerian speeds, while the accretors have Keplerian velocities  \citep{weidenschilling_1977}.
}

Shattering upon accretion may be a feature rather than a bug in the context of comet 81P/Wild-2, whose thermally processed, micron-sized particles have been described as ``chondrule and CAI-like fragments'' \citep{brownlee_etal_2012}. Perhaps mm-sized melt droplets strewn across the outer solar system by planetesimal collisions were broken up into micron-sized fragments upon being accreted by comet progenitors/Kuiper belt objects.
Low efficiencies of accretion might not be a problem either, insofar as micro-chondrules make up only a small fraction of the volume of the comet.

\subsection{Self-gravity}
\rev{Can chondrules produced from a collision
collapse under their own self-gravity? A necessary but not sufficient condition for gravitational collapse is that the local density exceed 
\begin{align} 
    \rho_{\rm Roche} &\approx \frac{3.5\Msun}{a^3} \nonumber \\
    &\approx 8 \times 10^{-8}\,\mathrm{g\,cm^{-3}}\left(\frac{a}{3\,\mathrm{ au}}\right)^{-3},
\end{align}
the minimum density a self-gravitating body must have to resist tidal disruption by the central star \citep[e.g.][]{chiang_youdin_2010}.
For our fiducial parameters, 
the mean density of the cloud of dust and chondrules just after it re-fills with nebular gas is $\rho_{\rm stall} + \rhon \sim 6 \rhon$, which falls short of $\rho_{\rm Roche}$ by
three orders of magnitude. Thus the cloud as a whole 
will eventually be sheared apart by solar tides and phase-mixed
with the rest of the nebula over a timescale
of order the orbital period, $\sim$$10^8$ s at 3 au.}

Might there be localized overdensities of particles
that self-gravitate? 
In their hydrodynamic 
simulation of a collision between $\sim$100-km
sized asteroids, \cite{stewart_lpsc3} found
clumps of debris having densities exceeding $\rho_{\rm Roche}$
and suggested they might 
collapse to form $\sim$10-km sized
planetesimals. These overdensities may be spurious if they arise from the unrealistic near-spherical collapse of their vapor plume (see end of section \ref{subsec:backfill}). Regardless, for a clump to self-gravitate it is not sufficient to satisfy the Roche criterion
against tidal disruption; the Jeans criterion
comparing self-gravity to gas pressure must also
be met. For a 
clump having particle density
$\rho_{\rm particle}$ and gas density
$\rho_{\rm gas}$ to be bound, it must be at least as large as the
Jeans length 
\begin{align}
    R_{\rm Jeans} &\sim \frac{c_{\rm neb}(1 + \rho_{\rm particle}/\rho_{\rm gas})^{-1/2} }{\sqrt{G(\rho_{\rm gas} + \rho_{\rm particle})}} \,  
    \label{eqn:Rjeans}
\end{align}
\citep{sekiya_1998, cuzzi_etal_2008, shi_chiang_2013}. 
Equation (\ref{eqn:Rjeans}) accounts
for how particles not only add to the gravity of the clump
(denominator), but also 
lower its effective sound speed from the
pure-gas value $c_{\rm neb}$ (numerator), 
in the limit where the gas-particle
mixture acts as a single
fluid on the clump free-fall time; this limit is
appropriate to chondrules which have short drag stopping times in the solar nebula. 
For parameters
inspired by the simulation
of \cite{stewart_lpsc3} ($\rho_{\rm gas} \sim 10^{-9}$ g/cm$^3$, $\rho_{\rm particle} \sim 10^{-6}$ g/cm$^3$, $c_{\rm neb} \sim 0.5$ km/s), we estimate $R_{\rm Jeans} \sim 4 \times 10^4$ km, which is $\sim$40 times larger than the clump sizes reported by these authors, precluding gravitational instability.
Turbulent motions only make it harder for
particle overdensities to become bound
\citep[e.g.][]{klahr_schreiber_2020b, klahr_schreiber_2020a}.

While gravitational instability in the immediate vicinity and aftermath of a vaporizing collision appears unviable, one could
wait for chondrules to settle to the disc midplane, where they could concentrate and perhaps eventually self-gravitate. The streaming instability \citep{youdin_goodman_2005} provides a route for particles to achieve super-Roche and super-Jeans densities \citep[e.g.][]{carrera_etal_2015, simon_etal_2017}.

\section{Summary and Discussion}
\label{sec:conclusions}

A theory for the formation of
chondrules should
explain how these mm-sized constituents
of the oldest known asteroids 
were melted and cooled, and how
they were collected with such efficiency as to fill $\gtrsim 50$\% of the volume
of meteorite parent bodies.
In this paper we have asked whether 
chondrites can form through
hypervelocity (vaporizing) collisions between
initially solid planetesimals, as suggested by
many workers (e.g. \citealt{campbell_etal_2002}; 
\citealt{krot_etal_2005}; \citealt{johnson_melosh_2015}; \citealt{johnson_etal_2016}; \citealt{fedkin_etal_2015}; \citealt{stewart_lpsc7}). 
We found that high-velocity collisions can
reproduce the thermal histories
of chondrules, \reva{but does not lend itself to understanding how they 
agglomerated into meteorite parent bodies}.

In a hypervelocity collision between two asteroids, a fraction of the colliding mass is vaporized. 
We have detailed the thermal and dynamical evolution of the initially hot, over-pressured vapor as it expands into the ambient nebula. 
The evolutionary stages of the vapor cloud include: (1) its initial free expansion and adiabatic cooling, (2) its condensation into dust grains which at first slow the cloud's cooling by releasing latent heat, and which later hasten cooling by radiating to space, (3) the cloud's deceleration due to loading by nebular gas, and (4) the cloud's eventual dissolution as nebular hydrogen backfills the near pressure-less cavity left by condensation. Against this evolving backdrop we have shown that solid/liquid particles entrained by the vapor
cloud can experience heating and cooling episodes consistent with those inferred for chondrules: heating to the point of melting for a period of order $10^2$ s, super-liquidus cooling at rates of 3000 K/hr or more, and sub-liquidus cooling at rates of 3000 K/hr to $< 100$ K/hr.

\reva{Cooling rates vary primarily with the total mass of the vapor cloud. As cloud masses increase from $10^{17}$ to $10^{20}$ g, sub-liquidus cooling rates decrease from 1000 to 100 K/hr. These cooling rates match those inferred experimentally for CB chondrules \citep{hewins_etal_2018}. For these same cloud masses, plume temperatures and pressures vary from $2500$ K and $10^{-2}$ bar on timescales of minutes, to $1500$ K and $10^{-6}$ bar over hours. These temperature and pressure ranges can reproduce the elemental abundances and zoning profiles of metal grains and chondrules in CB chondrites \citep{fedkin_etal_2015}. Although this agreement is encouraging, we have not shown that our time-dependent plumes vary slowly enough to host the equilibrium condensation sequences that \citet{fedkin_etal_2015} computed; moreover, their plume composition differs from ours, perhaps in significant ways.}

\revb{With ab initio simulations of vaporizing collisions still under development (cf. \citealt{stewart_lpsc7}), it remains unclear how efficiently mass can be converted into vapor (see also \citealt{carter_stewart_2020}, who estimated the frequency of vaporizing collisions among planetesimals in the primordial asteroid belt, but did not quantify the amount of vapor produced). If we co-opt the jetted melt fractions computed by \citet{johnson_melosh_2015} and \citet{wakita_etal_2017,wakita_etal_2021}, and assume that roughly 1\% of the mass in colliding bodies is converted into vapor, then 
the cloud masses of $10^{17}-10^{20}$ g that we find reproduce CB chondrules implicate colliding planetesimals with radii of 10--100 km. Planetesimals in this size range are typical of the solar system's minor body reservoirs \citep{morbidelli_etal_2009, sheppard_trujillo_2010}. For conventional, non-CB chondrules whose porphyritic textures imply slower cooling rates of 5--100 K/hr \citep{desch_connolly_2002, connolly_jones_review}, our model scalings point to clouds having masses  $\gtrsim 10^{22}$ g, or colliding bodies with radii $\gtrsim 500$ km. In some asteroid belt formation scenarios, many Ceres and Moon-sized bodies are thought to have populated the belt before being dynamically ejected \citep[e.g.][]{wetherill_1992, petit_etal_2001, o_brien_etal_2007, morbidelli_etal_2009}. }

\reva{\revb{If chondrules were ejected into the nebula by collisions, how did they agglomerate into a 
chondrite?} High-velocity collisions are a double-edged sword: while they generate enough heat to melt chondrules, they also implicate \reva{a population of} fast-moving planetesimals that cannot re-accrete chondrules efficiently and without damage. \reva{In the dynamically hot environment that we have envisioned, where non-circular velocities of large asteroids readily exceed $1$ km/s (corresponding to eccentricities and inclinations $> 0.05$),} gravitational focussing between chondrules and asteroids is negligible; an asteroid can accrete only those chondrules lying directly in its path, and not without shattering them on contact
(\citealt{wurm_etal_2005, teiser_etal_2009}).
\reva{Perhaps chondrules were accreted intact by especially small asteroids whose orbits were kept circular and co-planar by nebular gas drag.} 
Other chondrule concentration mechanisms include
the streaming instability \citep{youdin_goodman_2005, carrera_etal_2015, simon_etal_2017}, turbulent concentration (\citealt{hartlep_cuzzi_2020}), 
particle-aggregate sticking \citep{matsumoto_etal_2019},
and trapping in overpressured,
possibly self-gravitating gas rings
\citep[][]{tominaga_etal_2020}. A stringent test of any agglomeration theory is presented by the CB/CH chondrite Isheyevo, whose sedimentary laminations imply gentle, layer-by-layer accretion of size and mineral-sorted material \citep{garvie_etal_2017}. We have shown that the re-accretion/fallback scenario outlined by \citet{morris_etal_2015} does not pass this test.}

That accretion and shattering go hand in hand in high-velocity collision scenarios for chondrules might actually help to explain the ``chondrule fragments'' collected from the coma of comet 81P/Wild-2 \citep[e.g.][]{nakamura_etal_2008}. These micro-chondrules may be the broken, micron-sized remains of mm-sized solids that fragmented upon being accreted onto comet progenitors/Kuiper belt objects. Chondrule fragments might similarly coat the surfaces of asteroids --- they can be looked for in the \textit{Hayabusa2} and \textit{OSIRIS-REx} sample returns.

\section*{Acknowledgements}
\rev{We thank Sarah Stewart for an inspiring talk that motivated this work, Steve Desch for an insightful referee report that led to qualitative changes to our paper, and Jeffrey Fung for prompting us to consider the effects of the nebular headwind and sharing his numerical simulations of solid/gas interactions.} We thank Erik Asphaug, Bill Bottke,  Don Brownlee, Linda Elkins-Tanton, Sivan Ginzburg, Brandon Johnson, Philipp Kempski, Sasha Krot, Rixin Li, Tomoki Nakamura, Laura Schaefer, Shigeru Wakita, Ben Weiss, and Andrew Youdin for useful exchanges. We are also grateful to the many people we talked with over the years about chondrules, including Jay Melosh. This work used the \textsc{matplotlib} \citep{hunter_etal_2007} and \textsc{scipy} \citep{scipy_2020} packages, and was supported by NASA grant NNX15AD95G/NEXSS and Berkeley's Esper Larsen, Jr. fund. 

\section*{Data availability}
No new data were generated or analysed in support of this research.

\appendix




\bibliographystyle{mnras}
\bibliography{planets_nick} 




\bsp	
\label{lastpage}
\end{document}